\documentclass[aps,groupedaddress,prb,twocolumn,showpacs,superscriptaddress]{revtex4}
\usepackage{graphicx}
\usepackage{amsmath,amssymb,amsfonts}
\usepackage{natbib}

\newcommand{\be}{\begin{equation}}
\newcommand{\ee}{\end{equation}}

\begin{document}

\title{Quantum Markovian activated surface diffusion of
interacting adsorbates}

\author{R. Mart\'{\i}nez-Casado}
\email{ruth@imaff.cfmac.csic.es}

\affiliation{Instituto de F\'{\i}sica Fundamental,\\
Consejo Superior de Investigaciones Cient\'{\i}ficas,\\
Serrano 123, 28006 Madrid, Spain}

\author{A. S. Sanz}
\email{asanz@imaff.cfmac.csic.es}

\affiliation{Instituto de F\'{\i}sica Fundamental,\\
Consejo Superior de Investigaciones Cient\'{\i}ficas,\\
Serrano 123, 28006 Madrid, Spain}

\author{S. Miret-Art\'es}
\email{s.miret@imaff.cfmac.csic.es}

\affiliation{Instituto de F\'{\i}sica Fundamental,\\
Consejo Superior de Investigaciones Cient\'{\i}ficas,\\
Serrano 123, 28006 Madrid, Spain}

\date{\today}

\begin{abstract}
A quantum Markovian activated atom-surface diffusion model with
interacting adsorbates is proposed for the intermediate scattering
function, which is shown to be complex-valued and factorizable into
a classical-like and a quantum-mechanical factor.
Applications to the diffusion of Na atoms on flat (weakly corrugated)
and corrugated-Cu(001) surfaces at different coverages and surface
temperatures are analyzed.
Quantum effects are relevant to diffusion at low surface temperatures
and coverages even for relatively heavy particles, such as Na atoms,
where transport by tunneling is absent.
\end{abstract}

\pacs{68.35.Fx,05.10.Gg,68.43.Jk}

\maketitle


In 1954, van Hove\cite{vanhove} introduced the space-time correlation
function $G$ (a generalization of the well-known pair-distribution
function from the theory of liquids) as a tool to study the scattering
of probe particles off quantum systems consisting of ensembles of
interacting particles.
Within the Born approximation in scattering theory, the nature of the
scattered particles as well as the details of the interaction potential
are largely irrelevant.
Hence, following Lovesey,\cite{lovesey} the scattering processes with
interacting particles essentially reduce to a typical problem of
statistical mechanics.
The linear response function of the interacting particles, also known
as {\it dynamic structure factor} or {\it scattering law}, is then
related to the spontaneous-fluctuation spectrum of such particles
(measured from $G$) by the fluctuation-dissipation theorem and can be
expressed in terms of particle density-density correlation
functions.\cite{lovesey}
In general, $G$ is a complex-valued function, what can be considered
as a signature of the quantum nature of the problem.
The imaginary part of $G$ is important at small values of time (of the
order of $\hbar\beta$, with $\beta = 1/k_B T$), extending its range of
influence by decreasing the temperature.
This dynamical regime takes place when the thermal de Broglie
wavelength $\lambda_B = \hbar/\sqrt{2mk_BT}$ ($m$ is the adsorbate
mass) is of the order of or greater than the typical interparticle
distances.
The quantum system is then assumed to obey the fundamental condition of
stationarity and the scattering problem satisfies the detailed balance
principle, displaying the {\it recoil effect}.
Here, we study the quantum observable effects of Na-atom diffusion on
flat (weakly corrugated) and corrugated-Cu(001) surfaces probed by He
atoms at different values of the Na coverage ($\theta$) and the
surface temperature.
For simplicity, in our model only coupling to substrate phonons
(phonon friction) and not to low-lying electron-hole pair excitations
(electronic friction) is considered.
Nevertheless, as an extension of this stochastic model, the electronic
friction could be accounted for by simply adding it to the total
friction coefficient.
Moreover, diffusion by tunneling will not be considered in this work.

The observable magnitude in this type of scattering experiments is the
so-called {\it differential reflection coefficient}, which gives the
probability that the He atoms reach a certain solid angle $\Omega$ with
an energy exchange $\hbar\omega =E_f - E_i$ and wave vector transfer
parallel to the surface $\Delta {\bf K} = {\bf K}_f - {\bf K}_i$ after
probing the quantum system, which consists of an ensemble of
interacting Na atoms on the surface.
This magnitude reads as
\begin{equation}
 \frac{d^2  {\mathcal R} (\Delta {\bf K}, \omega)}{d\Omega d\omega}
   =  n_d {\mathcal F} S(\Delta {\bf K}, \omega) ,
 \label{eq1}
\end{equation}
where $n_d$ is the concentration of Na atoms on the surface,
${\mathcal F}$ is the {\it atomic form factor}, which depends on the
He-Na interaction potential, and $S(\Delta {\bf K},\omega)$ is the
dynamic structure factor or scattering law, which gives the line shape
and provides a complete information about the dynamics and structure of
the adsorbates.
Experimental information about long-distance correlations is obtained
from the dynamic structure factor for small values of $\Delta {\bf K}$,
while information on long-time correlations is provided at small energy
transfers, $\hbar \omega$.

The dynamic structure factor can be expressed as
\begin{eqnarray}
 S(\Delta {\bf K},\omega) & = & \frac{1}{2 \pi \hbar}
  \int \! \! \! \int e^{i (\Delta {\bf K} \cdot {\bf R} - \omega t)}
   \ \! G({\bf R},t) \ \! dt d{\bf R} \nonumber \\
& = & \frac{1}{2 \pi} \int e^{-i\omega t} \ \!
   I(\Delta{\bf K},t) \ \! dt  ,
 \label{eq2}
\end{eqnarray}
where ${\bf R}$ denotes the adatom position on the surface and
\be
 I(\Delta {\bf K},t) =
  \langle e^{- i \Delta {\bf K} \cdot {\bf R}(0)}
   e^{i \Delta {\bf K} \cdot {\bf R}(t)}  \rangle .
 \label{eq3}
\ee
In the Heisenberg representation, the trajectories ${\bf R}(t)$ are
replaced by linear position operators $\hat{{\bf R}}(t)$ and,
therefore, the brackets in Eq.~(\ref{eq3}) will denote the average
of the expectation value of the operators enclosed.
The so-called {\it intermediate scattering function},
$I(\Delta {\bf K},t)$, is the space Fourier transform
of the van Hove $G$-function.
Within the context of $^3$He spin-echo experiments, this function is
also known as {\it polarization},\cite{allison} its real and imaginary
parts being observable magnitudes [for instance, in Cs diffusion on a
corrugated Cu(001) surface\cite{jardine}].
An exact, direct calculation of $I(\Delta {\bf K},t)$ is difficult
to carry out due to the noncommutativity of the particle position
operators at different times.
Nonetheless, this calculation can be alternatively performed after
postulating\cite{chandrasekhar,vineyard} that the adatom position
operators $\hat{{\bf R}}(t)$ obey a standard Langevin equation (i.e.,
the associated stochastic dynamics is Markovian) and the product of
the two exponential operators in Eq.~(\ref{eq3}) can then be evaluated
according to a special case of the Baker-Hausdorff theorem
({\it disentangling theorem}), namely $e^{\hat{A}} e^{\hat{B}}
= e^{{\hat{A}}+{\hat{B}}} e^{[{\hat{A}}, {\hat{B}}]/2}$, which
only holds when the corresponding commutator is a c-number.
As shown below, this expression allows us to interpret
$I(\Delta {\bf K},t)$ as the product of a classical-like and a
quantum-mechanical intermediate scattering function associated
with the exponentials of $\hat{A}+\hat{B}$ and the commutator
$[\hat{A},\hat{B}]$, respectively.
Expressing now the thermal average implicit in Eq.~(\ref{eq3}) and the
corresponding position operators in terms of the system Hamiltonian,
and replacing $t$ by $t+i\hbar\beta$, the so-called {\it detailed
balance condition} reads as\cite{schofield}
\begin{equation}
 S(- \Delta {\bf K},- \omega) = e^{\hbar \beta \omega}
  S(\Delta {\bf K},\omega)  .
\end{equation}

Recently, it has been shown\cite{ruth1,ruth2} that results obtained
from a standard Langevin equation with two different non-correlated
noise sources compare fairly well with the experimental data available.
These noise sources are a Gaussian white noise accounting for the
surface friction and a white shot noise replacing the pairwise
interaction potential which simulates the adsorbate-adsorbate
collisions.
In this context, the double Markovian assumption holds because the
substrate excitation time scale is much shorter than the characteristic
times associated with the adatom motion (the maximum frequency of the
substrate excitation is around 20-30~meV and the characteristic
vibrational frequency of the adatom is about 4-6~meV).
Moreover, the time involved in a collision process is shorter than the
typical time between two consecutive collisions.
Thus, memory effects can be neglected. Within this framework,
called the {\it interacting single adsorbate} (ISA)
approximation, the total friction $\eta$ thus consists of the substrate
friction $\gamma$ and the {\it collisional friction} $\lambda$ (i.e.,
$\eta = \gamma + \lambda$).
The collisional friction can be related\cite{ruth3} to the surface
coverage $\theta$ as $\lambda = (6\rho\theta/a^2) \sqrt{k_B T/m}$,
where $a$ is the length of an assumed surface lattice with square unit
cells and $\rho$ is the effective radius of an adparticle.
Thus, the stochastic single-particle trajectories ${\bf R}(t)$ running
on the surface are assumed to obey the Langevin equation
\begin{equation}
 \ddot{\bf R}(t) = - \eta \dot{\bf R}(t) + {\bf F} ({\bf R} (t)) +
  \delta {\bf N}(t) ,
 \label{eq4}
\end{equation}
where ${\bf F} ({\bf R} (t))$ is the adiabatic force per mass unit
derived from the periodic surface interaction potential, $\delta
{\bf N}(t)$=$\delta {\bf N}_G (t)$+$\delta {\bf N}_S (t)$ is the
two-dimensional fluctuation of the total noise acting on the adparticle
(the $G$ and $S$ subscripts stand for Gaussian and shot, respectively).
In the Heisenberg representation,
Eq.~(\ref{eq4}) still hods, its formal solution being
\begin{eqnarray}
 \hat{\bf R} (t) & = & \hat{\bf R}_0 + \frac{\hat{\bf P}_0}{m \eta} \ \!
  \Phi (\eta t) \nonumber \\
 & + & \frac{1}{\eta} \int_0^t \Phi (\eta t - \eta t')
  \left[ \hat{\bf F} (\hat{\bf R} (t')) + \delta \hat{\bf N} (t')
    \right] dt' ,
 \label{eq5}
\end{eqnarray}
where $\hat{\bf R}_0$ and $\hat{\bf P}_0$ are the adparticle position
and momentum operators at $t=0$, respectively, and $\Phi(x)=1-e^{-x}$.
From Eqs.~(\ref{eq3}) and (\ref{eq5}), and considering the
disentangling theorem, $I(\Delta{\bf K},t)$ can be expressed as
\begin{equation}
 I (\Delta {\bf K},t) \simeq
  I_q (\Delta {\bf K},t) I_c (\Delta {\bf K},t) ,
 \label{eq6}
\end{equation}
which is a product of a quantum intermediate scattering
function, $I_q (\Delta{\bf K},t)$, and a classical-like one,
$I_c (\Delta {\bf K},t)$.
The quantum contribution, governed by the commutator, is the same for
any type of surface regardless of its relative corrugation.
For weakly corrugated surfaces, which can be assumed as flat, the
commutator between the position and momentum operators is a c-number,
and the relation (\ref{eq6}) is exact.
However, in general, the presence of the adiabatic force introduces an
additional commutator, $[\hat{\bf R}_0,\hat{\bf F}(\hat{\bf R}(t))] =
(i\hbar)\partial \hat{\bf F}(\hat{\bf R}(t))/\partial \hat{\bf P}_0$,
where the dependence of the adiabatic force on the initial state
$(\hat{\bf R}_0, \hat{\bf P}_0)$ is through $\hat{\bf R}(t)$.
Assuming a Markovian regime (fast memory loss on the initial
conditions), the previous commutator is going to be negligible.
Thus, in both cases, $I_q$ will reads as
\begin{equation}
 I_q (\Delta {\bf K},t) \simeq
  \exp \left[ \frac{i \hbar \Delta {\bf K}^2}{2 \eta m} \ \!
   \Phi(\eta t) \right] =
  \exp \left[ \frac{i E_r}{\hbar} \frac{\Phi(\eta t)}{\eta} \right] ,
 \label{eq7}
\end{equation}
where $E_r = \hbar^2 \Delta {\bf K}^2/2m$ is the {\it adsorbate recoil
energy}.
As expected, the argument of $I_q$ becomes less important as the
adparticle mass and the total friction increase (and, therefore, as
the coverage also increases).
Furthermore, for $\hbar = 0$, we recover the standard classical
function: $I = I_c$.
In (\ref{eq7}), the time-dependence arises from $\Phi(\eta t)$.
Thus, at short times ($\lesssim \hbar \beta$), we find $\Phi(\eta t)
\approx \eta t$ and the argument of $I_q$ becomes independent of the
total friction, increasing linearly with time.
On the contrary, in the long-time limit, this argument approaches a
constant phase.

\begin{figure}
 \includegraphics[width=8cm]{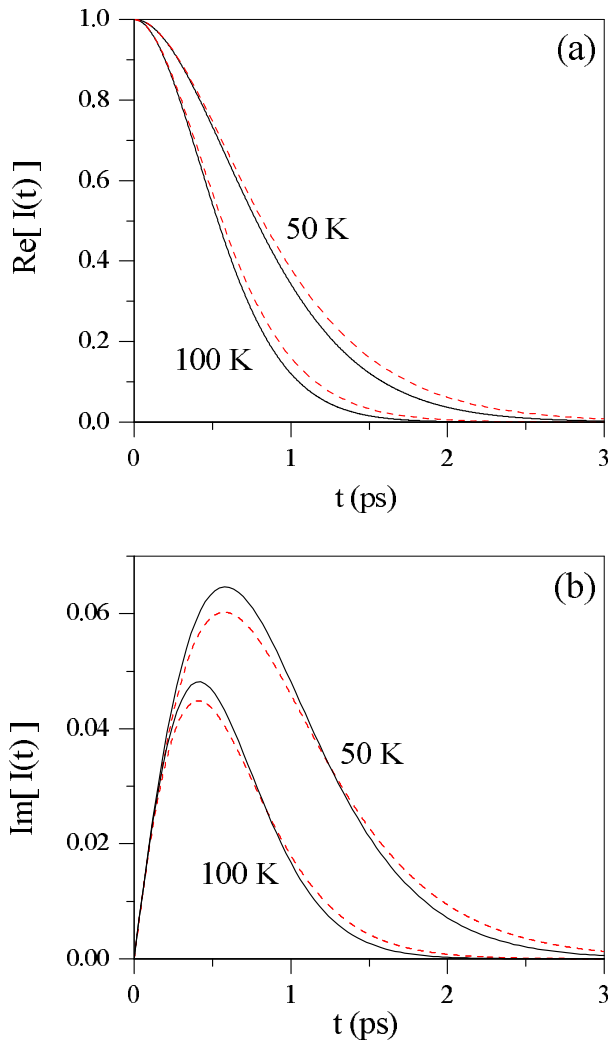}
 \caption{\label{fig1}
  (Color online)
  Quantum intermediate scattering function [Eq.~(\ref{eq9})] for Na
  diffusion on a flat surface at 50~K and 100~K: (a) real part and
  (b) imaginary part.
  Two coverages are considered: $\theta = 0.028$ (black solid line) and
  $\theta = 0.18$ (red/dark grey dashed line).}
\end{figure}

For flat or weakly corrugated surfaces, $I_c$ reads as
\begin{equation}
 I_c (\Delta {\bf K},t) = e^{- \chi^2 [ \eta t - \Phi(\eta t) ] } ,
 \label{eq8}
\end{equation}
where the so-called {\it shape parameter} $\chi$ is given by
$\chi^2 = \Delta {\bf K}^2 \langle {\bf v}^2_0 \rangle/\eta^2$ and,
therefore, the total intermediate scattering function (\ref{eq7}) can
be expressed as
\begin{equation}
 I (\Delta {\bf K},t) = e^{\alpha \chi^2} \,
e^{- \chi^2 [ \eta t + \alpha \Phi(\eta t) ] } ,
 \label{eq9}
\end{equation}
where $\alpha = 1 + i \hbar \eta / 2 k_B T$ if the thermal
square velocity is $\langle {\bf v}^2_0 \rangle = k_B T / m$.
Equation~(\ref{eq9}) is the generalization of the intermediate
scattering function for the quantum motion of interacting adsorbates
on a flat surface.
The dependence of this function on $\Delta {\bf K}^2$ through the shape
parameter is the same as in the classical theory.\cite{ruth3}
Note that the total intermediate scattering function issued from
Eqs.~(\ref{eq6})-(\ref{eq9}) is exact for Gaussian quantum processes
and no information about the velocity autocorrelation function is
needed.
However, classically the intermediate scattering function is
usually obtained from Doob's theorem, which states that the velocity
autocorrelation function for a Gaussian, Markovian stationary process
decays exponentially with time.\cite{risken}
Furthermore, two regimes are clearly distinguishable: free-diffusion
or ballistic and diffusive.
The former is dominant at very low times, $\eta t \ll 1$, while the
latter rules the dynamics at very long times, $\eta t \gg 1$.
In the diffusive regime, the mean square displacement,
$\langle {\bf R}^2 (t) \rangle$, is also linear with time, the
slope giving the diffusion coefficient according to Einstein's law,
$D = k_B T / m \eta$ (which insures that the adparticle velocity
distribution becomes Maxwellian asymptotically).
Therefore, since tunneling is absent, the quantum diffusion coefficient
follows Einstein's law as in the classical case.

In Fig.~\ref{fig1}, the real and imaginary parts of Eq.~(\ref{eq6})
for Na diffusion on a flat surface are plotted at two different surface
temperatures, 50 and 100~K, and two coverages, 0.028 and 0.18.
As can be clearly seen, the real part of $I(t)$ decreases faster with
temperature and slower with coverage.
On the other hand, the imaginary part of $I(t)$ displays maxima between
4-6\% of the corresponding real part, depending on the temperature.
It starts linearly with time and, after passing through a maximum,
decays smoothly to zero.
The corresponding quasielastic line shapes (around the zero energy exchange)
will then display narrowing with the coverage and broadening with the
surface temperature.
This behavior could be experimentally confirmed for those systems where
the diffusion barrier is smaller than the thermal energy $k_B T$.
For light particles, the imaginary part is expected to be much more
important keeping the same shape.

\begin{figure}
 \includegraphics[width=8cm]{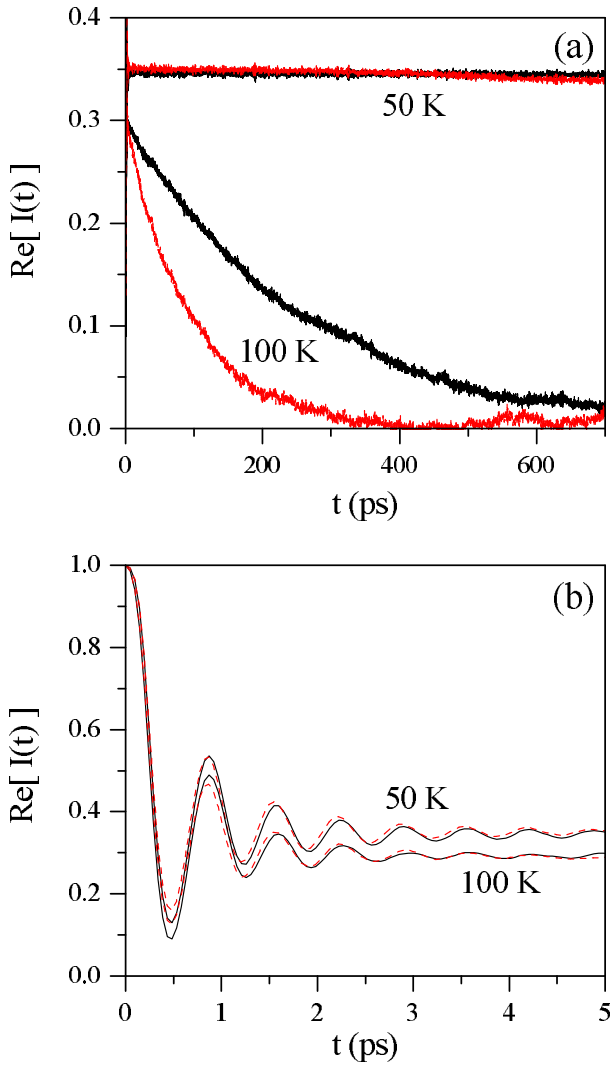}
 \caption{\label{fig2}
  (Color online)
  Real part (a) of the quantum intermediate scattering function
  [Eq.~(\ref{eq6})] for Na diffusion on a corrugated-Cu(001) surface
  at 50~K and 100~K.
  An enlargement at short time-scales is shown in part (b).
  Two coverages are considered: $\theta = 0.028$ (black solid line) and
  $\theta = 0.18$ (red/dark grey dashed line).}
\end{figure}

\begin{figure}
 \includegraphics[width=8cm]{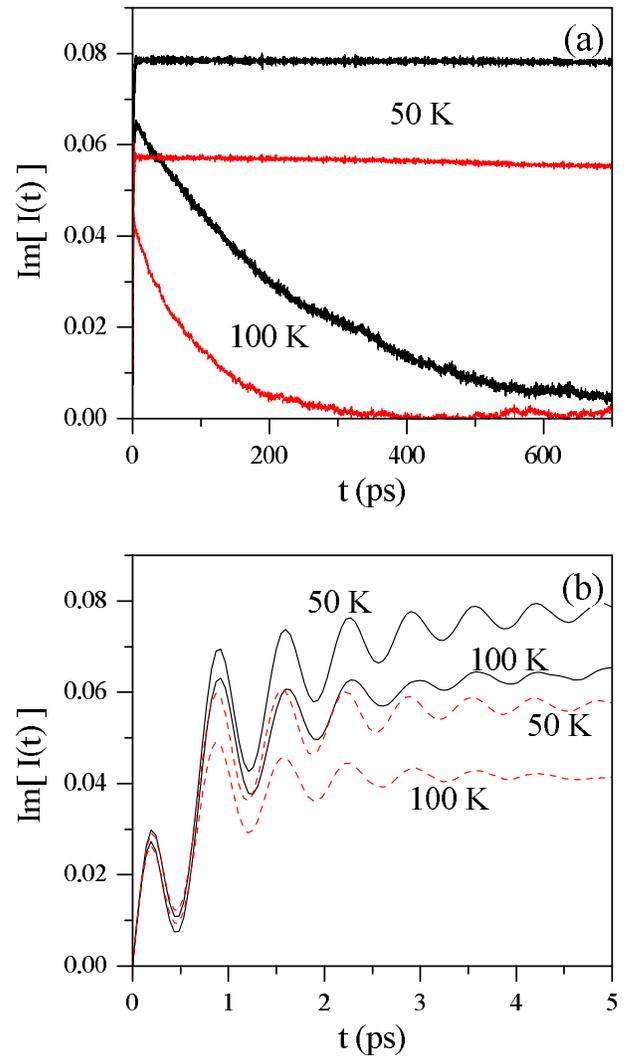}
 \caption{\label{fig3}
  (Color online)
  Imaginary part (a) of the quantum intermediate scattering function
  [Eq.~(\ref{eq6})] for Na diffusion on a corrugated-Cu(001) surface at
  50~K and 100~K.
  An enlargement at short time-scales is shown in part (b).
  Two coverages are considered: $\theta = 0.028$ (black solid line) and
  $\theta = 0.18$ (red/dark grey dashed line).}
\end{figure}

Let us now consider the case of nonzero corrugation.
For Na atoms, the pairwise interaction potential is repulsive and
the mean interparticle distance should be greater than $\lambda_B$
most of time.
Thus, the classical-like part of Eq.~(\ref{eq6}) could be replaced, at
a first approximation, by the classical counterpart.
Obviously, if diffusion is mediated by tunneling, this procedure is no
longer valid.
The error comes from small times, but since diffusion is a long-time
process, the influence on the quasielastic peak (wave-vector dependence)
and the quantum diffusion constant (Einstein's law) will be very small
for massive particles.
In Figs.~\ref{fig2} and \ref{fig3}, plots for Na diffusion on Cu(001)
at the same surface temperatures and coverages as in Fig.~\ref{fig1}
are shown.
The numerical values corresponding to $I_c$ have been obtained from
classical Langevin simulations in a nonseparable adsorbate-substrate
interaction potential.\cite{graham}
The global time behavior of the real and imaginary parts of the
intermediate scattering function [see Figs.~\ref{fig2}(a) and
\ref{fig3}(a), respectively] display important differences for high
and low temperatures.
At 50~K, the thermal energy is not enough to surmount the diffusion
barrier of the Cu(001) surface and adsorbates will remain for
relatively long times inside  potential wells.
On the contrary, at 100~K, the intermediate scattering function
decreases asymptotically to zero according to a more or less
exponential function,\cite{ruth3} as happens in a regime
characterized by surface diffusion.
The maxima displayed by the imaginary part are, again, around the same
percentage as before [see Fig.~\ref{fig3}(b)].
In principle, these imaginary parts should also be observable at
least at 100~K in spin-echo experiments.
With coverage, this time behavior means that the quasielastic
peaks, given by the scattering law, will undergo broadening, as
observed experimentally.\cite{toennies}
Quantum jump mechanisms can also be extracted from that peak in a
similar way to the classical procedure.\cite{ruth3}
The rapid oscillations displayed by the intermediate scattering
function at short times [see Figs.~\ref{fig2}(b) and \ref{fig3}(b)]
indicate bound motions inside the potential wells, which become less
pronounced as temperature increases.
These oscillations are associated with the lowest frequency mode
or frustrated translational mode.

\begin{figure}
 \includegraphics[width=8cm]{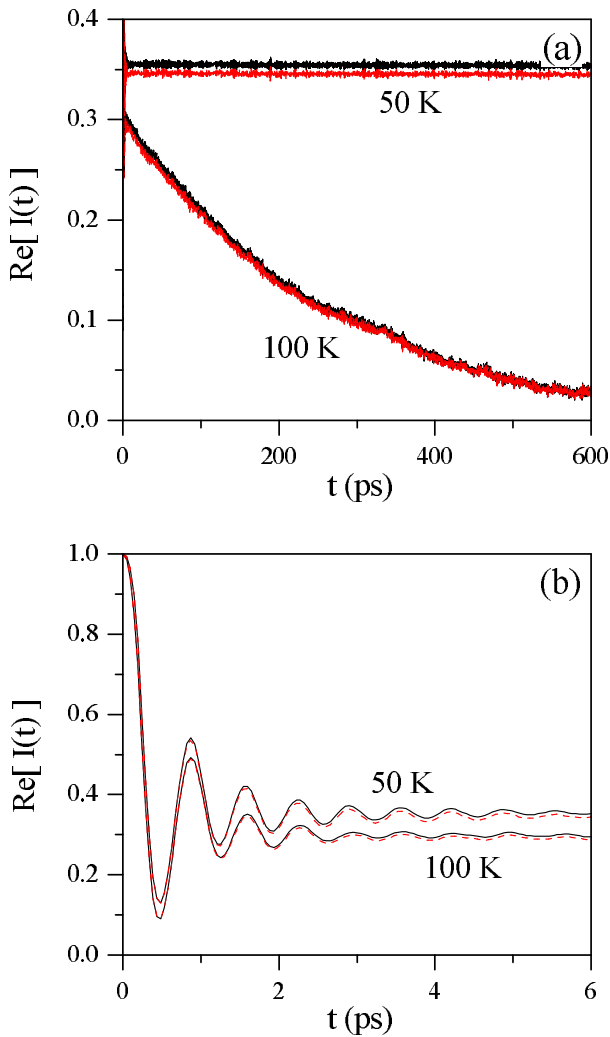}
 \caption{\label{fig4}
  (Color online)
  Classical intermediate scattering function for Na diffusion on a
  corrugated-Cu(001) surface at 50~K and 100~K (black solid lines)
  and the real part of its quantum-mechanical analog (\ref{eq6})
  (red/dark grey dashed lines).
  An enlargement at short time-scales is shown in part (b).
  The surface coverage considered here is 0.028.}
\end{figure}

In Fig.~\ref{fig4}, we show the effects of the quantum
correction in the diffusion process studied here at two different
surface temperatures for a coverage of 0.028.
To compare, both the classical intermediate scattering function and the
real part of its quantum analog are displayed in the figure.
As seen, although the Na atom is a relatively massive particle, at low
temperatures the {\it plateau} is lower for the quantum case.
This implies an initially (relatively) faster decay arising from the
strong influence of the quantum behavior at short time scales.
It is therefore the real part of the intermediate scattering function
what one should compare to the experiment rather than $I_c$, as
is usually done. Obviously, this effect will be less pronounced at
high coverages.

It is remarkable that, within the Markovian approach presented here,
the quantum intermediate scattering function, $I_q$, is independent
of the relative corrugation of the surface and, at short times, also
independent of the friction.
At low surface temperatures, the $I_q$ factor will be responsible
for a higher contribution of the imaginary part of $I$, given by
Eq.~(\ref{eq6}), modifying substantially the response in the diffusion
process.
Despite we have termed $I_c$ the classical-like intermediate scattering
function, strictly speaking it is not a classical magnitude, because it
involves position operators.
Nevertheless, for relatively heavy particles and at very long times
(diffusion time scales), operators can be replaced by variables, since
$\lambda_B$ is very small.
As far as we know, an exact quantum calculation for a corrugated
surface is not possible and some approximations have to be invoked,
e.g., the damped harmonic oscillator, which has been applied by some
of us\cite{david} within the same context.
Of course, other different, alternative theoretical approaches
can also be found in literature (see, for instance,
Ref.~\onlinecite{others}) within the single adsorbate approximation.
The theoretical formalism that we propose here should also be very
useful to avoid extrapolations at zero surface temperature when trying
to extract information about the frustrated translational mode.
Diffusion experiments at low temperatures are very difficult to
perform (or even unaffordable).
However, the type of theoretical calculations needed in this formalism
is easy to carry out and they would provide a simple manner to go to
very low temperatures with quite reliable results, thus allowing to
extract confident values of magnitudes such as friction coefficients
and oscillation frequencies.
By decreasing the surface temperature, quantum effects are extended
at higher values of time.
Going from 100~K to 50~K, the time where the quantum dynamics is
important increases from 0.07~ps to 0.15~ps.
The standard propagation time for diffusion is greater than 400~ps
as can be seen in Figs.~\ref{fig2} and \ref{fig3}.
In our opinion, the range of applicability of this quantum theory
should be around or below 10~K and with coverages up to 20\%.
Clearly, tunneling-mediated diffusion, where the isotopic effect and
the so-called crossover temperature have been first observed by Gomer
{\it et al.}\cite{gomer} is not accounted for by our model.
This is a very important aspect which deserves further consideration,
in particular, regarding the new observations of Zhu {\it et
al.}\cite{zhu} concerning the coverage dependence of tunneling
diffusion and the works by Ho and coworkers\cite{ho} and Sundell and
Wahnstr\"om.\cite{sundell}

Finally, it is well known that the broadening is due to the increase of
coverage.
As we have shown in Ref.~\onlinecite{ruth3}, better agreement with the
experimental results is found with the ISA model (classical theory)
than with previous calculations where the repulsive lateral interaction
is taken into account in Langevin molecular dynamics simulations.
This lead us to conclude that a stochastic description of the
broadening should be good enough to describe it since the statistical
limit (central limit theorem) in the number of collisions should be
reached in the very long time propagation describing the diffusion
process.
This conclusion is still valid in this quantum theory since the
adparticle is massive and the small quantum effects are mainly observed
in the short time limit.
What this quantum theory has showed is the complex character of the
intermediate scattering function whose real and imaginary parts are
observable,\cite{jardine} even for massive adparticles (Cs atoms).

We would like to thank the Cambridge Surface Physics group for very
interesting and stimulating discussions.
This work has been supported by the Ministerio de Ciencia e
Innovaci\'on (Spain) under project with reference FIS2007-62006.
R.\ Mart\'{\i}nez-Casado and A.S.\ Sanz acknowledge the Consejo
Superior de Investigaciones Cient\'{\i}ficas for a predoctoral contract
and a JAE-Doc contract, respectively.


\end{document}